\documentclass[a4paper,12pt]{article}

\usepackage[utf8]{inputenc}
\usepackage[T1]{fontenc}
\usepackage {graphicx}
\usepackage{amsmath}
\usepackage{amssymb}

\usepackage{setspace}

\begin{document}
\title{Evolutionary food web model based on body masses gives realistic networks with permanent species turnover}
\author{K.T. Allhoff \thanks{Corresponding author: allhoff@fkp.tu-darmstadt.de 
| Institute for Condensed Matter Physics, Technical University of Darmstadt, Germany} ,
D. Ritterskamp, 
\thanks{Institute for Chemistry and Biology of the Marine Environment, Carl von Ossietzky University of Oldenburg, Germany} ,\\ 
B.C. Rall, 
\thanks{German Centre for Integrative Biodiversity Research (iDiv) Halle-Jena-Leipzig, Germany;
Institute of Ecology, Friedrich Schiller University Jena, Germany;
Netherlands Institute of Ecology (NIOO-KNAW), Wageningen, The Netherlands;
J.F. Blumenbach Institute of Zoology and Anthropology, Georg-August-University G\"ottingen, Germany 
} ,
B. Drossel,  
\thanks{Institute for Condensed Matter Physics, Technical University of Darmstadt, Germany} ,
C. Guill
\thanks{Institute for Biodiversity and Ecosystem Dynamics, University of Amsterdam, The Netherlands}}


\maketitle


\begin{abstract}
The networks of predator-prey interactions in ecological systems are remarkably complex, 
but nevertheless surprisingly stable in terms of long term persistence of the system as a whole. 
In order to understand the mechanism driving the complexity and stability of such food webs, 
we developed an eco-evolutionary model in which new species emerge as modifications of existing ones and dynamic ecological interactions determine which species are viable. 
The food-web structure thereby emerges from the dynamical interplay between speciation and trophic interactions. 
The proposed model is less abstract than earlier evolutionary food web models in the sense that all three evolving traits have a clear biological meaning, 
namely the average body mass of the individuals, the preferred prey body mass, and the width of their potential prey body mass spectrum. 
We observed networks with a wide range of sizes and structures and high similarity to natural food webs. 
The model networks exhibit a continuous species turnover, but massive extinction waves that affect more than $50 \%$ of the network are not observed.    
\end{abstract}


\clearpage
\doublespacing


\section{Introduction}
Classical models addressing the structure and stability of food webs are based on stochastic algorithms that produce structural patterns similar to empirically measured food webs \cite{Drossel2005}, 
such as the niche model \cite{Nische2000} or the cascade model \cite{Cohen1985}. 
A more recent approach is to use the empirically found allometries of body size and foraging behaviour of individual consumers to predict the links between species on a more biological basis \cite{Petchey2008}.  

However, real food webs are not produced by a generative algorithm, but have been shaped by their evolutionary history and show an ongoing species turnover. 
New species in a food web occur by immigration and speciation, and species vanish due to extinction.
Currently, the world faces one of the largest extinction waves ever, which is thought to be caused by anthropogenic drivers such as climate change and land use \cite{Barnosky2011}. 
Even without human interference or other catastrophic causes, and apart from evolutionary suicide due to runaway selection \cite{Parvinen2005}, biological extinctions occur due to intrinsic processes, 
i.e., the dynamic trophic and competitive interactions among species \cite{Riede2011b, Binzer2011}. 
The stability of food webs in terms of resistance to extinction waves after a perturbation (such as the removal or addition of a species), 
thus also depends on the network structure of these interactions between the species \cite{May1972, Otto2007}, and conversely the network structure results from species extinctions and additions.
Understanding the interplay of food web structure and stability has therefore been identified as one of the most important questions in ecology \cite{May1999}.

Over the last decade, several models were introduced that include evolutionary rules on a longer time scale, in addition to population dynamics on shorter time scales: 
The former enables new species to enter the system, whereas the latter determines which species are viable and which go extinct. 
The newly emerging species can be modelled and interpreted either as invaders from another, not explicitly considered region or as ``mutants'' of existing species.
The emerging network structures evolve in a self-organising manner, giving rise to complex, species-rich communities even when starting from initial networks with very few species \cite{Drossel2005}.

A particularly simple and often cited evolutionary food web model was introduced in 2005 by Loeuille and Loreau \cite{LL2005} and subsequently modified by several other authors \cite{Ingram2009, Brannstrom2011, Allhoff2013}. 
Each species is characterised by its body mass, which is the only evolving trait. 
Feeding and competition interactions are determined via differences in body mass. 
The fact that body mass is an ecologically interpretable trait makes the results from this model easily comparable to empirical data. 
This major advantage has been pointed out in the review on large community-evolution models by Br\"annstr\"om et al. \cite{Brannstrom2012}. 
The evolutionary process in this model generates large networks that show an almost static behaviour, with clearly defined niches all of which are and remain occupied. 
Even if a newly emerging species is slightly better adapted to the resources and therefore displaces a species of similar body mass, 
it has the same feeding preferences and hence the same function in the food web, leading to a very low species turnover without secondary extinctions \cite{Allhoff2013}. 
The network structure is robust with respect to various changes in the population dynamics rules, indicating that some simple, robust mechanism structuring these food webs is at work \cite{LL2005, Allhoff2013}. 

Complex networks with a less rigid structure emerge in the evolutionary version of the niche model \cite{Guill2008}. 
The model allows for the evolution of three traits instead of just one, namely the niche value, the centre and the width of the feeding range. 
Other authors describe a species in a more abstract way by a vector of many traits, as implemented in the matching model \cite{Rossberg2006, Rossberg2008} and in the webworld model \cite{Drossel2001, Drossel2004}. 
Recently, also several individual-based models for evolving food webs were introduced \cite{Bell2007, Yamaguchi2011, Takahashi2013}. 
The emergence of complex food webs in these models is highly nontrivial. 
Some past attempts to set up an evolutionary model lead to repeated network collapse instead of persisting complex networks \cite{Takahashi2011}. 
Other attempts lead to trivial network structures, like simple food chains  in the evolving niche model \cite{Guill2008} or a single trophic level in the webworld model \cite{Drossel2004}. 
In both models, adaptive foraging was required in order to obtain more complex networks.  

Allhoff and Drossel \cite{Allhoff2013} suggested that an evolutionary food web model has to fulfil two conditions to be able to generate diverse and complex networks. 
First, it should allow for the evolution of more traits in addition to body mass in order to generate several possible survival strategies like for example specialists and omnivores. 
This idea is consistent with results from a recent empirical study by Rall et al \cite{Rall2011}, who found that predators of similar body mass differ significantly in their feeding preferences. 
Second, the evolution of each trait has to be restricted in order to prevent unrealistic trends, for example towards extremely small or large body masses or towards extremely broad or narrow feeding ranges. 
In this context, the stabilising effect of adaptive foraging in previous models could be explained by the fact that a predator can focus on its most profitable prey without losing adaptation to other prey.

In this paper, we propose a new evolutionary food web model that includes the restriction of trait evolution in a more direct way. 
Similarly to the evolutionary niche model \cite{Guill2008} and supported by empirical data regarding the body-mass ratios of predator-prey pairs \cite{Brose2006a, Riede2011}, 
we characterise a species by three traits with clear biological meaning: 
its own body mass (which determines its metabolic rates), its preferred prey body mass, and the width of its potential prey body mass spectrum. 
The evolutionary rules in our model confine the traits within certain boundaries, without the requirement to include adaptive foraging. 

The model most similar to our model is the one by Loeuille and Loreau \cite{LL2005}. 
It also has body mass as a key trait and a similar concept for setting the feeding preferences. 
Our model differs from the model by Loeuille and Loreau in the number of traits that characterize a species (3 instead of 1), 
the functional response (Beddington-deAngelis instead of linear), 
the competition rules (based on link overlap instead of body mass differences), 
the possibility of cannibalism and loops (included only in our model) and the resource dynamics. 
Moreover, we consider body mass ratios instead of body mass differences 
so that the body masses in our model spread over several orders of magnitude instead of only one. 
The bio-energetics of the species in our model follow well documented allometric scaling relationships \cite{Brown2004}, 
leading to networks with realistic body-mass scaling relations that can be tested directly against empirical data. 

We demonstrate the capabilities of our model by evaluating 18 common food web properties and compare them to a data set of 51 empirical food webs from a large variety of different ecosystems. 
We further use the well-known evolutionary model by Loeuille and Loreau \cite{LL2005} as a benchmark to assess the quality of the predictions of our model. 
In principle, both models are able to produce diverse networks. 
However, we obtain a higher variability in the feeding preferences and survival strategies and therefore more realistic values for the corresponding network properties. 
Moreover, while the network structures of Loeuille and Loreau are static, species turnover and extinction avalanches occur naturally in our model. 


\section{The Model}
The model includes fast ecological processes (population dynamics), 
which determine whether a species is viable in a given environment that is created by the other species, 
and slow evolutionary processes (speciation events), which add new species and enable the network to grow and produce a self-organised structure.
A species $i$ is characterised by its body mass, $m_i$, the centre of its feeding range, $f_i$, and the width of its feeding range, $s_i$. 
These traits determine the feeding interactions in the community (see Fig. 1) and thereby the population dynamics. 
A summary of all model parameters and variables is given in tab. 1.


\subsection*{Population dynamics}
The population dynamics follows the multi-species generalisation of the bioenergetics approach by Yodzis and Innes \cite{Yodzis1992,Brose2006}. 
The rates of change of the biomass densities $B_i$ of the populations are given by
\begin{equation}
\dot{B_0} = G_0 B_0 -  \hspace{-1.0em} \sum_{j=\text{consumers}} \hspace{-1.2em} g_{j0}B_j\\
\end{equation}
for the external resource (species 0) and
\begin{equation}
\dot{B_i} = \hspace{-1.0em} \sum_{j=\text{resources}} \hspace{-1.2em} e_jg_{ij}B_i -\hspace{-1.0em} \sum_{j=\text{consumers}} \hspace{-1.2em} g_{ji}B_j\, - \,x_i B_i
\end{equation}
for consumer species. 
$G_0 = R(1-B_0/K)$ is the logistic growth rate of the external resource, $e_j$ is the efficiency with which biomass of species $j$ can be assimilated by its consumers, 
$g_{ij}$ is the mass-specific rate with which species $i$ consumes species $j$, and $x_i$ is $i$'s mass-specific respiration rate. 
The mass-specific consumption rate is given by
\begin{equation}
g_{ij} = \frac{1}{m_i}\frac{a_{ij} B_j}{1+\sum_{k=\text{res.}}h_ia_{ik}B_k + \sum_{l=\text{comp.}}c_{il}B_l}\,,
\end{equation}  
where 
\begin{equation}
 a_{ij} = a_i\cdot N_{ij} = a_i\cdot \frac{1}{s_i\sqrt{2\pi}} \cdot  \exp\left[-\frac{(\log_{10}f_i - \log_{10}m_j)^2}{2s_i^2}\right]
\end{equation}
is the rate of successful attacks of species $i$ on individuals of species $j$, with the Gaussian feeding kernel $N_{ij}$ given in Fig. 1.  
The parameter $h_i$ is the handling time of species $i$ for one unit of prey biomass, 
and $c_{il}$ quantifies interference competition among predators $i$ and $l$ \cite{Beddington1975, DeAngelis1975, Skalski2001}. 
It depends on their similarity, as measured by the overlap $I_{il} = \int N_{ij}\cdot N_{lj} d(\log_{10} m_j)$ of their feeding kernels, via
\begin{equation}
 c_{il} = c_{\text{food}} \cdot \frac{I_{il}}{I_{ii}} \;\;\;\;\text{for}\;\;\;\; i\neq l.
\end{equation}
The normalisation of the competition with $I_{ii}$ was proposed by Scheffer et al. \cite{Scheffer2006} and accounts for the fact that the competition matrix is not symmetric. 
More specialised species exert a higher competition pressure than species with broad feeding ranges. 
The overlap $I_{il}$  is similar to the niche overlap discussed by May \cite{May1974}.

We assume that interference competition is significantly higher within a species than between different species, e.g. due to territorial or mating behaviour. 
To account for this, we introduce an intraspecific competition parameter $c_{\text{intra}}$ and set $c_{ii} = c_{\text{food}} + c_{\text{intra}}$.


\subsection*{Speciation events}
Each simulation starts with a single ancestor species with body mass $m_1=100$ and feeding parameters $f_1=1$ and $s_1=1$, 
which is thus feeding on the external resource with its maximum attack rate. 
The initial biomass densities are $B_0=K=100$ for the resource and $B_1 = m_1\cdot \epsilon = 2\cdot 10^{-2}$ for the ancestor species. 
The parameter $\epsilon$ is the extinction threshold, i.e., the minimum population density required for a population to survive. 
At each unit time step, species below this extinction threshold get removed from the system. 

A speciation event occurs with probability $\omega = 0.0001$ per unit time. 
This is so rare that the system is typically close to a fixed point before the next mutation occurs. 
Then, one of the currently existing species (but not the external resource) is chosen randomly as parent species $i$ for a ``mutant'' species $j$. 
Thus, every species has the same probability $\omega / S$ to ``mutate'', where $S$ is the number of currently viable species. 
The logarithm of the mutant's body mass, $\log_{10}(m_j)$, is chosen randomly from the interval $[\log_{10}(0.5m_i), \log_{10}(2m_i)]$, 
meaning that the body masses of parent and mutant species differ at most by a factor of 2. 
The mutant's initial biomass density is set to $B_j = m_j\cdot \epsilon$ and is taken from the parent species. 

The mutant's feeding traits $f_j$ and $s_j$ are independent of the parent species. 
The logarithm of the feeding centre, $\log_{10} f_j$, is drawn randomly from the interval $\left[\left(\log_{10}(m_j)-3\right) , \left(\log_{10}(m_j)- 0.5 \right)  \right]$, 
meaning that the preferred prey body mass is 3 to 1000 times smaller than the consumer's body mass, and following the results from Brose et al. \cite{Brose2006a}. 
The width of the feeding range, $s_j$, is drawn randomly from the interval $[0.5, 1.5]$. 
A small value of $s_j$ corresponds to a more specialised consumer, while a large value of $s_j$ characterises a consumer with a broad feeding range and lower attack rates. 
A combination of large preferred prey mass $f_j$ and a wide feeding range enables a consumer to prey on species with a larger body mass than its own. 
This enables the emergence of cannibalism and feeding loops. 
The fixed intervals keep the evolving traits in reasonable ranges and prevent unrealistic trends, following the results by Allhoff and Drossel \cite{Allhoff2013}. 

When testing the robustness of the model predictions with respect to the model details, we used alternative rules, 
where the probability for choosing a parent species is proportional to its biomass (similar to the model by Loeuille and Loreau \cite{LL2005}) 
or to its inverse generation time $m_i^{-1/4}$ so that the mutation rate is proportional to the reproduction rate. 
Furthermore, we tested Gaussian distributions of mutant body masses around the parent with a standard deviation between 0.09 and 1. 
We used a cutoff at two standard deviations resulting in a maximum body mass factor between parent and daughter species between $10^{2\cdot 0.09}\approx 1.5$ and $10^{2\cdot 1}\approx 100$. 
The former describes local speciation events, whereas the latter describes species invasions from not explicitly modelled regions. 
We also compared the results to simulations where the mutants body mass is drawn randomly from the interval $[10^{-0.5},10^6]$.
Finally, with a similar approach, we also included heredity into the feeding parameters $s_i$ and $f_i$ 
by combining Gaussian distributions around the parent's traits with the above given mutation intervals. 



\section{Methods}
The computer code for our simulations was written in C. 
We used the Runge-Kutta-Fehlberg algorithm provided by the GNU Scientific library \cite{GSL} for the numerical integration of the differential equations. 
Simulations were run for $5\cdot 10^8$ time units. 
For comparison, the generation time of the initial ancestor species with body size $m_1 = 100$ is of the order of $\frac{1}{x_1}=\frac{100^{0.25}}{0.314}\approx 10$ time units. 

The competition parameters $c_{\text{food}}$ and $c_{\text{intra}}$ have a strong effect on the diversity of the emerging food webs. 
To obtain the network variability observed in nature, we performed computer simulations with all four combinations of $c_{\text{food}} = 0.6$ or $0.8$ and $c_{\text{intra}} = 1.4$ or $1.8$. 
The time series of these simulations are shown in the online supplementary material. 
From each simulation run, we collected 80 food webs obtained after every $5\cdot10^6$ time units from $t=10^8$ to $t=5\cdot 10^8$, resulting in a total of 320 different networks. 
Due to the initial build-up of the network, the first $10^8$ time units were not taken into account. 

The structure of the emerging food webs is compared both to food webs produced with the model by Loeuille and Loreau \cite{LL2005} and to empirical food webs. 
For the empirical data, we re-evaluated 51 of the 65 food webs from different ecosystem types analysed by Riede et al. \cite{Riede2010} 
for which we had body-mass data for all species in the network (for the complete list see online supplementary material).
For the model by Loeuille and Loreau, we evaluated the final network structures obtained with 75 combinations of different parameter values.  
Due to the static network structure, we could not obtain different networks from one evolutionary simulation.
The niche width of \cite{LL2005} was set to $nw=\frac{s^2}{d}=0.5, 1.0, 1.5, 2.0, 2.5$ and the competition strength to $\alpha_0=0.1, 0.2, 0.3, 0.4, 0.5$, similar to the original work. 
To get networks of comparable size we decreased the competition range, $\beta=0.025, 0.05, 0.075$. 

Both models use Gaussian feeding kernels with in principle infinite width to describe the feeding interactions, meaning that each species can prey on every other species. 
Thus, for analysis, very weak links have to be cut off in order to obtain meaningful network structures. 
In our networks, we removed all links that contribute less than $75\%$ of the average link to the total resources of a consumer. 
This criterion is weaker than it might seem, because most of the links of a predator are very weak, and so is the average link strength. 
Our cutoff measure depends on both the attack rate and the prey's biomass density. 
It thereby mimics unavoidable sampling limits in empirical food-web studies. 
For the networks produced by the algorithm of Loeuille and Loreau we used the cutoff criterion of the original work 
and removed all links with an attack rate less than $15\%$ of the respective predator's potential maximum attack rate, disregarding the prey's biomass density. 
Since the value of the cutoff criterion is to some extent arbitrary, we report its effects on the predicted network properties in the online supplementary material. 
There we also show results obtained for the model by Loeuille and Loreau with our cutoff rule.

\section{Results}
\subsection*{A typical simulation run}
A typical simulation run with the competition parameters $c_{\text{intra}}=1.4$ and $c_{\text{food}}=0.8$ is shown in Fig. 2. 
After an initial period of strong diversification, the system reaches a size of approximately $60$ species (panel (a)) 
on 3 to 4 trophic levels above the resource (panel (c)).  
The species form clusters of similar body masses, as shown in panel (b). 
New predator and prey species emerge preferentially within these clusters: 
A prey species in a cluster experiences less predation pressure due to the saturation of the functional response of the predator, 
and the predation input of a predator is larger if its feeding preferences match such a cluster. 
Therefore, we observe a trend towards strong specialisation on these clusters, resulting in the following network structure. 
Species in the first cluster have a body mass of approximately $10^1$, specialise on the resource and represent most of the second trophic level. 
Species in the second cluster with a body mass of approximately $10^2-10^3$ feed either on the resource ($\text{TL}\approx2$) or on the first cluster ($\text{TL}\approx3$). 
Species in the top cluster with a body mass greater than $10^3$ specialise either on the first or on the second cluster and therefore have intermediate trophic positions ($3\leq\text{TL}\leq4$). 
Some species have even higher trophic positions due to cannibalism and loops.

The initial build-up of the network continues until the species in the top cluster are close to the extinction threshold.
Once all clusters have emerged, the system shows a continuous turnover of species. 
We suppose the following turnover mechanism. 
Mutants with very few predators can occur occasionally if their body mass is between two clusters and if the other species are specialised on the clusters. 
If such a mutant has viable feeding parameters, it can grow a large population and displace many other species at once, potentially even causing secondary extinctions. 
Examples for such extinction events are visible at $t\approx 2 \cdot 10^8$ and $t\approx 4.3\cdot 10^8$. 
After an extinction event, the network rearranges, and temporally also species with broader feeding ranges appear, 
before the trend towards specialisation followed by an extinction event starts again.


\subsection*{Network evaluation and comparison}
We compared 320 networks from our model with 51 empirical networks and 75 networks from the model by Loeuille and Loreau \cite{LL2005}, see Fig. 3.
Panels (a)-(c) show the distributions of body masses of all three data sets.
The observed peaks in our simulated data correspond to the body mass clusters mentioned before. 
The distance between the peak maxima is determined by the upper boundary of the mutation interval of the feeding centre. 
Single empirical food webs show a similar peak pattern (not shown). 
In contrast, the body mass distribution of the model by Loeuille and Loreau looks blurred, due to our choice of the niche width $nw=\frac{s^2}{d}$. 
With smaller values of the feeding range $s$, the network structure is strongly layered and clusters of body masses that are multiples of the feeding distance $d$ occur, 
where each species feeds on those in the cluster below and is prey to those in the cluster above \cite{Allhoff2013}.
We also observed that because we based the network structure on predator-prey body-mass ratios instead of body-mass differences, 
the resulting community-size spectra from our model follow empirical observations and theoretical predictions more closely than those from the model by Loeuille and Loreau, as shown in the online supplementary material. 

Panels (d)-(f) show the distributions of trophic levels of all three data sets. 
Here, we use the short-weighted instead of the flow-based trophic level. 
This allows for better comparison with the empirical data for which the population sizes are often not available. 
The comparison between the two models reveals the main difference between the two different cutoff rules. 
A link with intermediate attack rate to a small prey population represents only a small proportion of the predator's diet, 
and is therefore neglected when using our cutoff threshold ($75\%$ of the average link). 
However, it is not recognised as a weak link with the cutoff rule by Loeuille and Loreau ($15\%$ of the maximum attack rate). 
On the other hand, a link with small attack rate to a big prey population (especially to the resource) is deleted in their model. 
Thus, trophic levels are overestimated, whereas our model with our cutoff rule results in a quite realistic distribution. 

Both models have difficulties reproducing the empirical distributions of generality (number of prey species) and vulnerability (number of predators), 
which are much broader than the distributions produced by the models (panels (g)-(l)). 
For the model by Loeuille and Loreau, the distribution results from the fact that the species in the model feed only on prey with smaller body masses. 
The situation is similar to the cascade model \cite{Cohen1985}, which also constrains predators to feed only on prey with a lower rank. 
Consequently, both generality and vulnerability cannot be larger than twice the average number of links per species. 
In our new model, the distribution of the vulnerability shows two humps. 
The first hump contains the carnivores in the higher trophic levels that feed on herbivores or on other carnivores. 
They have a high generality and a small vulnerability.
The second hump contains the herbivores that feed on the resource. 
They are prey to many other species and hence have a high vulnerability. 

We ascribe the differences between the models and the empirical distributions to the fact that both models have only one resource, 
which means that all herbivores feed on the same resource, whereas in empirical networks herbivores can have more than one resource. 
Furthermore, both models ignore the within-species body-mass distribution by assigning to each species a precise value of the body mass. 
This also narrows down the range of body sizes a species can feed on or is vulnerable to.


By analysing the $320$ networks from the 4 simulations separately (see Fig. 4), we found two trends concerning the network size (panel (a)):
First, the stronger the intraspecific competition $c_{\text{intra}}$, the smaller are the population sizes and the more populations can survive on the same amount of energy provided by the resource. 
Second, the stronger the competition for food $c_{\text{food}}$ is, the sooner species can displace others resulting in rather small networks with fast evolutionary species turnover. 

Both models are able to produce networks of realistic sizes, but tend to overestimate the number of links per species (panel (b)) and hence the connectance (panel (c)). 
The effect is much larger in the model by Loeuille and Loreau due to their original cutoff rule. 
This also explains the high fraction of omnivores and the low fraction of top and herbivorous species (panels (d)-(f)), 
as well as the high values of the number of chains and the clustering coefficient (panel (o) and (p)) and the small value of the characteristic path length (panel (r)).  
In the online supplementary material, we show that the model by Loeuille and Loreau provides more realistic predictions when using our cutoff rule.

Both models fail to reproduce the maximum similarity (panel (q)), due to the same reasons that also lead to the narrow distributions of generality and vulnerability.
For the remaining panels, the model by Loeuille and Loreau performs worse than our model regardless which cutoff rule is used. 
For example, the short-weighted trophic levels (panel (j)-(l)) are not only overestimated due to the cutoff rule, but also reflect the regular network structure. 
As mentioned above, these networks are layer-like structures, where each cluster represents one trophic level. 
Since all clusters accommodate a similar number of species instead of heaving more species on lower levels like in our model, the mean trophic level is overestimated. 
Moreover, the model does not include cannibalism (panel (m)) and loops (panel (n)), for which our model provides good predictions. 

Due to the evolution of three instead of one trait, we obtain more diverse network structures than Loeuille and Loreau. 
We observe a higher standard deviation of the generality, the vulnerability and the linkedness (panel (g)-(i)), reflecting different feeding preferences and survival strategies.


\subsection*{Robustness of the results against variations of the evolutionary rules}

In order to ensure that our findings are no artefacts of the specific choice of evolutionary rules, 
we tested the robustness of our results against the changes outlined at the end of the model section. 
We found that making mutation probabilities dependent on biomass or body mass influences the time dependency of the network development 
but leaves the averaged network properties, like the total network size, the distribution of body masses and the fraction of species or biomasses per trophic level, mostly unchanged. 
Also the trend towards strong specialisation with subsequent extinctions still occurs in these variants. 

When changing the degree to which the parent's body mass is inherited by the mutant, the main effect was that species turnover became slower with stronger inheritance. 
In this case it is less likely that mutants with body masses between two clusters occur, which have few predators and cause extinction avalanches. 
The probability for such mutants increases with a decreasing degree of inheritance, which is consistent with our oberservation that the body mass clusters appear to be blurred in case of a very low degree of inheritance. 
The same is true for randomly chosen body masses. However, we still obtain large, complex networks.

If the parent species $i$ and the mutant $j$ have similar feeding centres, $f_i\approx f_j$, the initial build-up of different trophic levels and their recovery after an extinction avalanche is also slowed down. 
With very strong inheritance of the feeding centre, all species will focus on the resource 
and no mutant emerges with a feeding centre matching the first body mass cluster, leading to trivial structures with only one trophic level. 
If parent and mutant have a similar degree of specialisation, $s_i\approx s_j$, all species exert and experience a similar competition pressure. 
Thus, instead of one species displacing another, both populations stay small and hence more populations per trophic level can survive. 
However, small or intermediate degrees of inheritance in the feeding traits leave the network characteristics again mostly unchanged. 
The situation is different, when either the feeding range or the feeding centre is chosen from an interval around the parent's trait without any body mass dependent constraint. 
In consistency with the predictions of Allhoff and Drossel \cite{Allhoff2013}, these variants lead to unrealistic trends and trivial instead of complex network structures.

\section{Discussion}
We introduced a new evolutionary food web model where the feeding links are based on body mass, and where species differ by body mass, feeding centre, and feeding range. 
By iterating population dynamics and speciation events for a sufficiently long time, we obtained complex networks, which show a high degree of commonality with empirical food webs. 
The new model is able to produce more realistic and more diverse network structures than the model by Loeuille and Loreau \cite{LL2005}. 

Both models use a very similar approach of Gaussian feeding kernels to determine the interactions between the species, which by construction leads to perfectly interval networks. 
Following the results of Stouffer et al. \cite{Stouffer2006}, we assume this to be a reasonable approximation. 
In contrast to the model by Loeuille and Loreau, the new model allows for cannibalism and loops, since the feeding range can extend to body masses larger than that of the predator. 
The species in our model can have different feeding preferences and survival strategies, due to the larger number of evolving traits in our model. 
This leads to a higher variability in network characteristics such as linkedness, generality and vulnerability, even though natural variability is still larger, 
which we ascribe to the facts that our model has only one basal resource and no body-size structure within species. 
We showed that an appropriate choice of the cutoff rule for weak links is essential for obtaining realistic results for connectance and trophic structure.

The increased number of evolving traits compared to the model by Loeuille and Loreau has also a large effect on the evolutionary trends. 
The networks show an ongoing species turnover and are subject to constant restructuring. 
The species in our model form body mass clusters and the evolutionary process is characterised by a trend towards increased specialisation on these clusters. 
Similar specialisation trends have also been observed in other studies \cite{Guill2008, Allhoff2013}. 
We assume the following explanation for the continuous species turnover. 
The evolved specialists gradually replace less efficient species with broader feeding ranges that cover also the gaps between the body mass clusters. 
Those broad ranged species have the role of keystone species that stabilise the networks against the occurrence of large extinction avalanches \cite{Power1996, Ekloef2006}. 
In the absence of control by such predators, new mutants (or invaders) can find niches between two clusters with very little predation pressure, 
where they can grow to high abundance and cause extinction avalanches propagating from lower to higher trophic levels. 
After such extinction events, the empty niches can be reoccupied also by species with broader feeding ranges, before the speciation process starts again. 

This corresponds to the results of Binzer et al. \cite{Binzer2011}, who identified specialised species on high trophic levels to be prone to secondary extinctions, 
and to the results of Rossberg \cite{Rossberg2013}, who suggested a very similar turnover mechanism for the results of his model. 
In consistence with the described mechanism, also Mellard and Ballantyne \cite{Mellard2014} reported that co-evolution of species does not necessarily lead to high levels of resilience for the ecosystem as a whole. 
However, such a turnover mechanism is missing in the model by Loeuille and Loreau. 
There, a displaced species is always replaced by a new species of a very similar body mass. 
And since the body mass is the only evolving trait, the new species has automatically the same predators and the same prey, 
excluding the possibility of secondary extinctions or major changes in the network structure \cite{Allhoff2013}. 
The same is true for the model version by Brannström et al., which led to evolutionary equilibria, 
where no more mutants are able to invade the system \cite{Brannstrom2011}. 
Ingram et al. reported that also their model extension with evolving feeding ranges, but with fixed predator-prey body mass ratios, 
tends to reach dynamically stable configurations with little structural change \cite{Ingram2009}. 

However, real ecosystems do show extinction events of different sizes, and their distribution evaluated over geological times resembles a power law \cite{Raup1986}. 
For this reason, it has been suggested that ecosystems show self-organised criticality (SOC) \cite{Sneppen1995}, 
which means that the intrinsic dynamics of the systems is responsible for the power-law size distribution of extinctions. 
However, the question remains open due to sparse and ambiguous data \cite{Newman2003, Drossel2001a}.
Some previous evolutionary food web models, for example the evolutionary niche model \cite{Guill2008}, exhibit SOC, 
whereas other models like the webworld model \cite{Drossel2001} or the model by Loeuille and Loreau \cite{LL2005} do not. 
The size distribution of extinction avalanches in our model is a power law with an exponent around $4$ (not shown). 
Because of its steepness, this power law covers only approximately one decade, meaning that extinction events of more than 10 species are extremely rare. 
This is not the type of SOC required to explain the large extinction events in earth history, where up to 90 percent of all species went extinct. 
Regarding the time span a species is present in the system, our model is consistent with paleobiological data concerning the fact 
that higher trophic level species stay in the system for a shorter time span than lower level species \cite{Newman2003}, 
although it should be mentioned that the exact distribution of these time spans in our model depends on the relation between a species' body mass and its mutation probability. 

The evolutionary rules implemented in our model are simplified and to some extent artificial. 
To make sure that our results do not depend on these simplified rules, we tested several variations concerning the mutation and inheritance rules. 
Our general finding is that minor changes in the evolutionary algorithm have only minor effects on the results. 
The overall mechanism with a trend towards specialisation followed by an extinction event as explained above is robust to changes in the evolutionary rules. 
Also the time averaged network structures remain mostly unchanged. 
However, the typical time period for a specialisation-extinction cycle can change with extinction events being triggered sooner or later. 

The fact that our networks show realistic patterns concerning many common food web properties suggests 
that our model provides a valuable tool to discuss urgent topics in ecological research. 
For example, the allometric equations are extendable by temperature terms (e.g. \cite{Vasseur2005, Binzer2012, Norberg2012, Stegen2011}). 
This approach would allow to model how warming might change evolution and extinction waves, in order to discuss current global change questions. 

Another idea would be to address habitat loss and habitat fragmentation as a prominent example of an external driver of extinction events \cite{Hagen2012, Gonzalez2011}. 
Recently, various approaches have been made to study the influence of the spatial environment on the food web composition and stability. 
If space has the structure of discrete habitats, these food webs can be interpreted as ``networks of networks'' \cite{Amarasekare2008, Leibold2004}.
However, most of the studies on such metacommunities so far focus on spatial aspects under the assumption that the species composition is static, 
although it has been emphasised that combining the spatial and the evolutionary perspective is essential for a better understanding of ecosystems \cite{Logue2011, Urban2008, Loeuille2008}. 
Recently, Allhoff et al. studied a spatial version of the model by Loeuille and Loreau \cite{Allhoff2014}. 
However, their findings were associated with the applied competition rules and the remarkable stability of the original model, 
highlighting the assumption that a more dynamic species turnover as in our new model would lead to a better understanding of the interplay between evolving food web structure and spatial structure.


\singlespacing
\small
\newpage

\addcontentsline{toc}{section}{Acknowledgements}
\section*{Acknowledgements}
This work was supported by the DFG under contract number Dr300/12-1 and 13-1. 
C.G. is supported by the Leopoldina Fellowship Program under contract number LPDS 2012-07. 
We thank Markus Schiffhauer and Jannis Weigend, who analysed several model variants concerning the mutation and heredity rules in the context of their bachelor theses. 
Moreover, we thank Christoph Digel and Jens Riede for providing empirical data and Sebastian Plitzko and Bernd Blasius for very helpful discussions.  

\addcontentsline{toc}{section}{Author contribution statement}
\section*{Author contribution statement}
All authors designed the model. 
K.T.A. performed the simulations. 
K.T.A. and C.G. analysed the results. 
K.T.A. wrote the manuscript with minor contributions from the other authors. 
All authors reviewed the manuscript. 

\addcontentsline{toc}{section}{Additional Information}
\section*{Additional information}
The authors declare no competing financial interests.



\newpage
\begin{table}[ht]
\centering \begin{tabular}{lll}
\hline 
parameter & meaning\\
\hline
resource  & \\
\hline 
$m_{0}=1$ 	&  body mass\\
$R=1$ 		& maximum mass-specific growth rate\\
$K=100$   	& carrying capacity\\
$B_{0}$ 	& biomass density\\
\hline 
species $i$ &  \\
\hline 
$m_{i}$ 	&  body mass\\
$f_{i}$ 	&  centre of feeding range\\
$s_{i}$ 	&  standard deviation of feeding range\\
$B_{i}$ 	&  biomass density\\
\hline 
population dynamics & \\
\hline
$e_j=0.85 \;(0.45)$  &  assimilation efficiency for animal (plant) resources\\
$g_{ij}$ 	&   functional response of predator $i$ on prey $j$\\
$a_{ij}$ 	& attack rate of predator $i$ on prey $j$\\
$a_{i}=1 \cdot m_i^{0.75}$ & attack rate parameter\\
$h_{i}=0.398 \cdot m_i^{-0.75}$  &  handling time of predator $i$\\
$c_{il}$ 	&  competition on species $i$ from species $l$\\
$c_{\text{food}}$ 	& competition parameter for food\\
$c_{\text{intra}}$ 	&  intraspecific competition parameter\\
$x_{i}=0.314 \cdot m_i^{-0.25}$  & respiration rate of species $i$\\
\hline 
evolutionary rules & & \\
\hline
$\omega=10^{-4}$  & mutation probability \\
$\epsilon =\frac{2}{10^{4}}$  & initial population density of a new species and extinction threshold\tabularnewline
\hline
\end{tabular}\caption{A summary of all model parameters. 
The values of the population parameters are based on \cite{Yodzis1992}. 
If no value is given for a parameter, it is variable.}
\end{table}

\newpage

\begin{figure}[ht]
  \centering
  \includegraphics[width=0.5\textwidth]{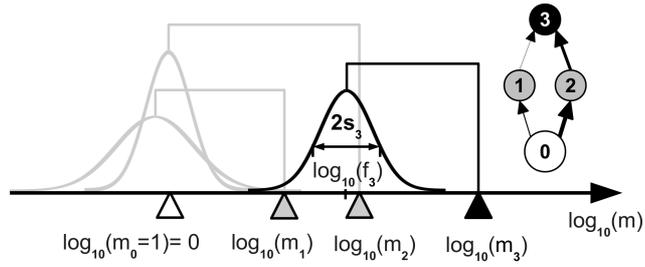}
  \caption{Model illustration using 4 species. Species $3$ (black triangle) is characterised by
its body mass $m_3$, the centre of its feeding range $f_3$, and the width of its feeding range $s_3$. 
The Gaussian function (black curve) describes its attack rate kernel $N_{3j}$ on potential prey species. 
Here, species $3$ feeds on species $2$ and $1$ (grey triangles) with a high resp. low attack rate. 
Species $1$ and $2$ are consumers of the external resource, represented as species 0 with a body mass $m_0=1$ (white triangle). 
Also shown is the corresponding network graph.}
  \label{fig:traits}
\end{figure}

\begin{figure}[ht]
 \centering
  \includegraphics[width=0.5\textwidth]{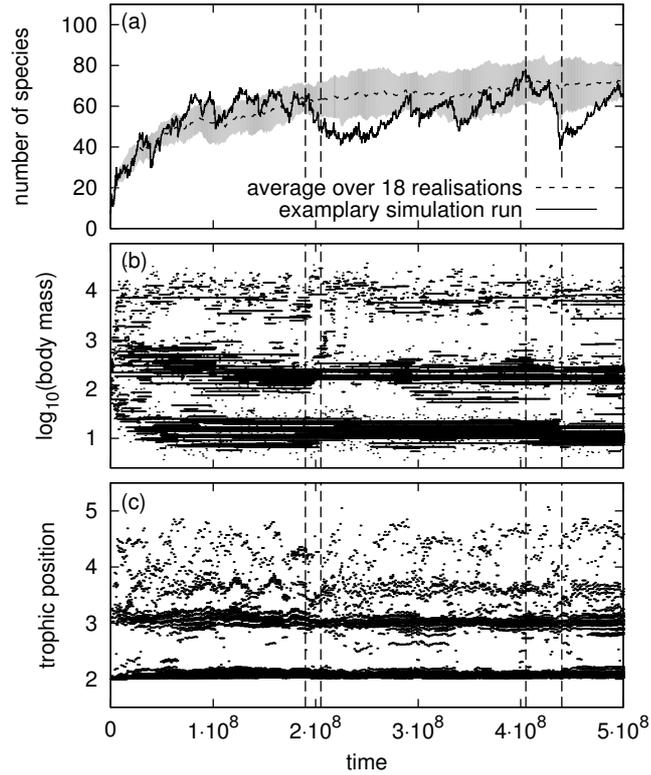}
 \caption{Network size, body masses and flow-based trophic positions \cite{Williams2004a} of all species 
 occurring during one exemplary simulation run with competition parameters $c_{\text{intra}}=1.4$ and $c_{\text{food}}=0.8$. 
 Panel (a) also shows the average network size and its standard deviation for 18 simulations with identical parameters but different random numbers. 
 Body masses and trophic positions were plotted at every 25th mutation event.  
 Network visualisations for the time points indicated by vertical lines are shown in the online supplementary material.}
 \label{fig:exrun1}
\end{figure}

\begin{figure}[ht]
 \centering
  \includegraphics[width=0.5\textwidth]{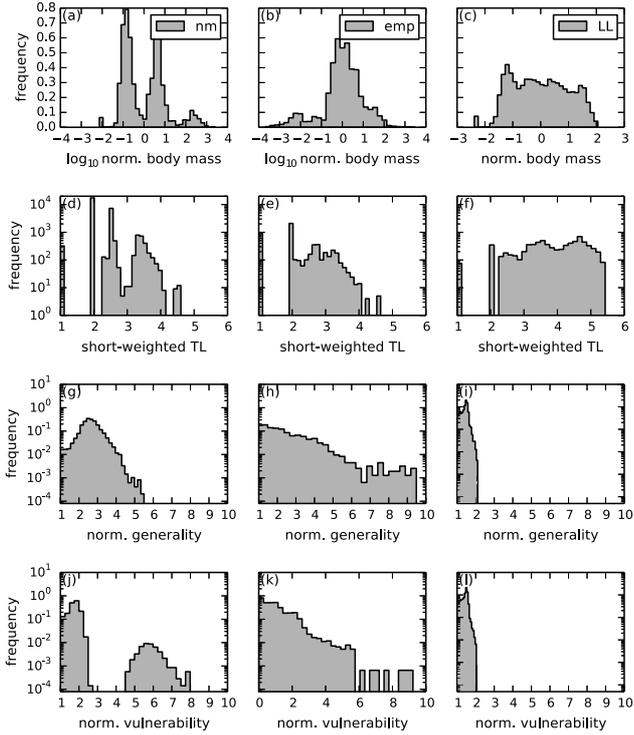}
 \caption{Frequency distributions of body masses and short-weighted trophic level \cite{Williams2004a}, 
 as well as the distributions of generality (number of prey species) and vulnerability (number of predators). 
The latter two are normalised by the average number of links per species. 
\textbf{nm}: $320$ networks from $4$ simulations 
of our new model with all four combinations of $c_{\text{food}} = 0.6$ or $0.8$ and $c_{\text{intra}} = 1.4$ or $1.8$. 
\textbf{emp}: Average over 51 empirical food webs.
\textbf{LL}: Average over 75 simulations of the model by Loeuille and Loreau \cite{LL2005}. 
Note that panel (c) shows absolute body masses, since in this model all body masses are in the same order of magnitude. 
See \emph{Methods} for more information.}
 \label{fig:distributions}
\end{figure} 

\begin{figure}[ht]
 \centering
  \includegraphics[width=\textwidth]{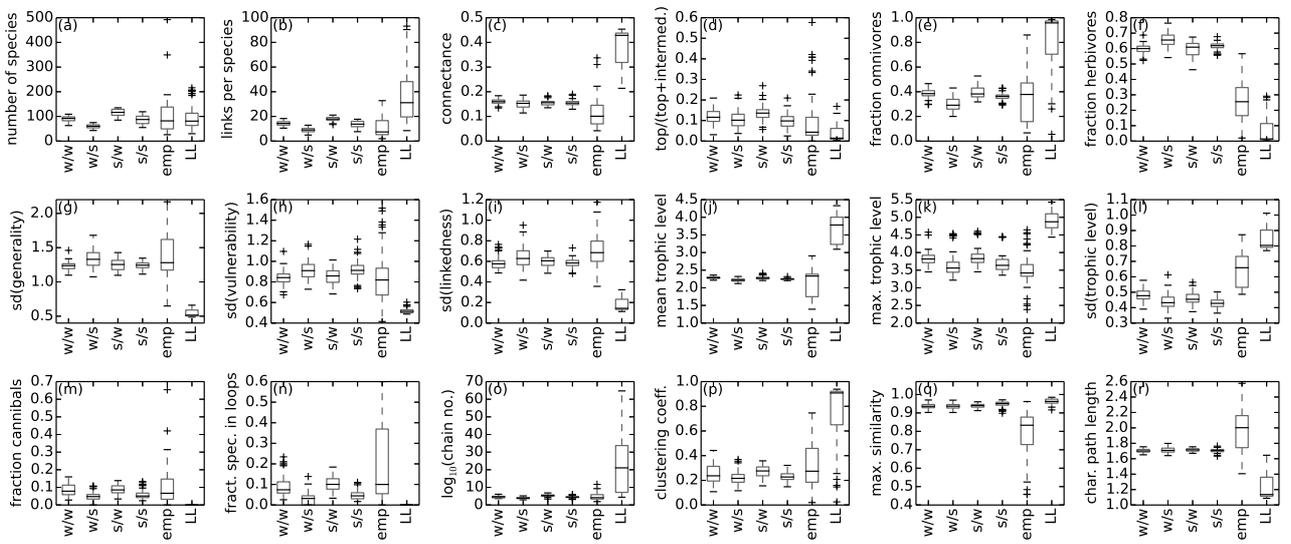}
 \caption{Network properties of four realisations with different values of the competition parameters. 
\textbf{w/w}: Weak competition, $c_{\text{intra}}=1.4$ / $c_{\text{food}}=0.6$.
\textbf{w/s}: Weak intraspecific competition and strong competition for food, $c_{\text{intra}}=1.4$ / $c_{\text{food}}=0.8$.
\textbf{s/w}: Strong intraspecific competition and weak competition for food, $c_{\text{intra}}=1.6$ / $c_{\text{food}}=0.6$. 
\textbf{s/s}: Strong competition, $c_{\text{intra}}=1.6$ / $c_{\text{food}}=0.8$. 
\textbf{emp}: Average over 51 empirical food webs.
\textbf{LL}: Average over 75 simulations of the model by Loeuille and Loreau \cite{LL2005}. 
See \emph{Methods} for more information.
Details on the calculation of these network characteristics can be found in the online supplementary material.}
 \label{fig:boxplots}
\end{figure}

\end{document}